\PassOptionsToPackage{bookmarks=false}{hyperref}
\documentclass[sn-nature]{sn-jnl}

\usepackage{graphicx}%
\usepackage{multirow}%
\usepackage{amsmath,amssymb,amsfonts}%
\usepackage{amsthm}%
\usepackage{mathrsfs}%
\usepackage[title]{appendix}%
\usepackage{xcolor}%
\usepackage{textcomp}%
\usepackage{manyfoot}%
\usepackage{booktabs}%
\usepackage{algorithm}%
\usepackage{algorithmicx}%
\usepackage{algpseudocode}%
\usepackage{listings}%
\usepackage{color,soul}
\usepackage{float}  
\usepackage[T1]{fontenc}
\usepackage{lineno}
\usepackage[none]{hyphenat}
\usepackage{setspace}
\usepackage{geometry}
\usepackage{subfiles}
\unnumbered

\begin{document}
\title[Engineering and exploiting self-driven domain wall motion in ferrimagnets for neuromorphic computing applications]{Engineering and exploiting self-driven domain wall motion in ferrimagnets for neuromorphic computing applications}


\author*[1,2]{\fnm{Jeffrey A.} \sur{Brock}}\email{jeffrey.brock@psi.ch}

\author[1,2]{\fnm{Aleksandr} \sur{Kurenkov}}

\author*[1,2]{\fnm{Ale\v{s}} \sur{Hrabec}}\email{ales.hrabec@psi.ch}

\author[1,2]{\fnm{Laura J.} \sur{Heyderman}}

\affil[1]{\orgdiv{Laboratory for Mesoscopic Systems}, \orgname{Department of Materials, ETH Zurich}, \orgaddress{\postcode{8093} \city{Zurich}, \country{Switzerland}}}

\affil[2]{\orgname{PSI Center for Neutron and Muon Sciences}, \orgaddress{\postcode{5232} \city{Villigen PSI}, \country{Switzerland}}}


\abstract{\begin{spacing}{1.5}Magnetic domain wall motion has recently garnered significant interest as a physical mechanism to enable energy-efficient, next-generation brain-inspired computing architectures. However, realizing all behaviors required for neuromorphic computing within standard material systems remains a significant challenge, as these functionalities often rely on competing interactions. Here, we demonstrate how spontaneous domain wall motion in response to locally engineered lateral exchange coupling in transition metal–rare earth ferrimagnets can be leveraged to achieve numerous neuromorphic computing functionalities in devices with minimal complexity. Through experiments and micromagnetic simulations, we show how tuning the feature size, material composition, and chiral interaction strength controls the speed of self-driven domain wall motion. When integrated with spin-orbit torque, this control gives rise to behaviors essential for neuromorphic computing, including leaky integration and passive resetting of artificial neuron potential. These results establish locally engineered ferrimagnets as a tunable, scalable, and straightforward platform for domain wall-based computing architectures.\end{spacing}} 




\maketitle 
In an effort to realize more energy-efficient approaches to computing, the past several years have seen significant research invested toward creating analogues of biological neural networks in condensed matter systems. Regardless of the physical property used to mimic neuron potentiation, an artificial neuron must simultaneously exhibit several capabilities, including (i) integration of input stimuli, (ii) leaking of potential between stimuli, and (iii) resetting back to an initial state after firing upon reaching a threshold potential \cite{ghosh2009spiking,izhikevich2003simple,brette2005adaptive,zhu2025thermally,wang2024compact}. Spintronic systems are particularly interesting in this context, with tunable functionalities that can provide rich nonlinear dynamics and stochasticity while maintaining a high endurance~\cite{grollier2020neuromorphic,zhou2021prospect,marrows2024neuromorphic,chen2023spintronic}. Among spintronic platforms for neuromorphic computing, magnetic domain walls (DWs) are exceptionally promising due to their ability to be moved and detected using electric currents. Additionally, the non-volatility of DWs facilitates in-memory computing, avoiding the energy and time inefficiency associated with separation of memory and processing in traditional von Neumann architectures \cite{siddiqui2019magnetic}. 

To relate magnetic DW motion to neuromorphic computing, analogies can be drawn between the behaviors of DWs and biological neurons. Namely, the position of a DW is analogous to the potential energy of a neuron, the accumulated DW displacement in response to a series of external stimuli emulates the neuronal integration, and neuron firing corresponds to the DW reaching a specific position \cite{hassan2018magnetic,Sengupta_2018, can,siddiqui2019magnetic}. To mimic biological neurons, the integration should be leaky, meaning that, without an input signal, the DW retreats with time \cite{jaiswal2020neural,brigner2019graded,mah2021domain,mah2023leakage}. Additionally, a passive reset of the DW to its original state after firing is necessary \cite{das2022self,azam2020voltage,siddiqui2019magnetic}. Given the diverse functionalities required of DW-based artificial neurons, designing materials and devices hosting competing interactions to enable all these behaviors in one device remains challenging \cite{hassan2018magnetic,fan2015stt,wang2023spintronic,liu2024domain}.

In parallel to the growing interest in using magnetic DWs for neuromorphic computing, transition metal-rare earth (TM-RE) ferrimagnetic alloys have regained interest for their exceptional magnetization dynamics useful for applications, including extraordinarily fast and efficient domain wall motion \cite{Caretta2018,blasing2018exchange,kim_ferrimagnet}. Ferrimagnets also generate low stray fields because of their low net magnetization, enabling denser device integration compared to ferromagnets \cite{finley2020spintronics}. Furthermore, the ease with which the magnetically dominant sublattice of TM-RE ferrimagnets can be locally modified using focused ion or electron beams, composition gradients, oxygen plasma, or laser annealing has enabled new magnetic behaviors \cite{Frackowiak2020,Frackowiak2021,LeCraw1974,Hrabec2011,Hansen1973,Lee2023,Krupinski2021,Riddiford2024,Ma2025}, including spontaneous DW motion in the absence of external stimuli \cite{Liu2023}. 

Here, we show that spontaneous DW motion arising from local control of the dominant magnetic sublattice in ferrimagnets enables precise tuning of the leak and reset behaviors necessary to construct artificial neurons, and several strategies to modulate these behaviors. We then integrate this control of spontaneous DW motion with spin–orbit torque to build neuronal structures that exhibit leaky signal integration and passive reset once the signal stops. Our results show how the fast and efficient DW motion in TM–RE ferrimagnets, together with precise local patterning to control the key parameters, provide a straightforward-to-implement platform for neuromorphic computing. 
\pagebreak

\begin{figure*}[htb!]
    \centering
    \includegraphics[width=0.9\textwidth]{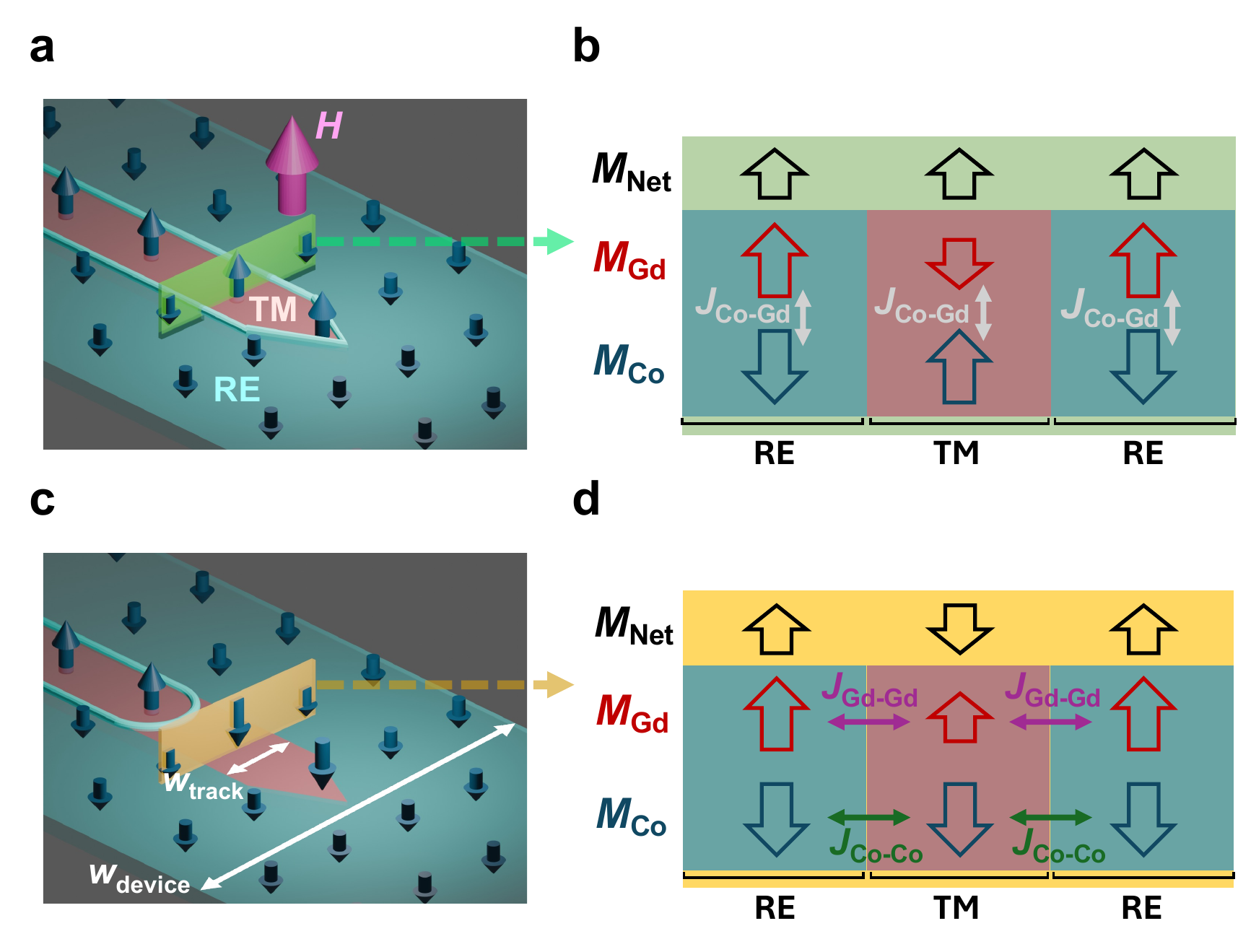}
    \caption{\textbf{Engineered self-driven DW motion in TM-RE CoGd ferrimagnet tracks.} \textbf{(a)}~Schematic illustration of a TM-dominant track (pink inner area) within an as-grown RE-dominant device (teal outer area) created with DWLA. In an applied magnetic field \textit{H} (magenta arrow), the net magnetizations of the TM- and RE-dominant regions $M_{\mathrm{net}}$ align parallel to \textit{H}, while still satisfying the antiferromagnetic Co-Gd exchange interaction $J_{\mathrm{Co-Gd}})$ as highlighted in the cross-sectional view of the magnetic configuration across the green plane in \textbf{(b)}. The DW between the TM- and RE-dominant regions is indicated by the light blue boundary in (a). \textbf{(c)} Schematic illustration of the domain configuration after \textit{H} is removed, indicating the track and device width ($w_{\mathrm{track}}$ and $w_{\mathrm{device}}$, respectively). \textbf{(d)}~Cross-sectional view of the magnetic configuration across the yellow plane in (c), demonstrating that, in the absence of \textit{H}, there is a reversal in the net magnetization of the TM-dominant region to minimize the ferromagnetic Co-Co and Gd-Gd exchange energies ($J_{\mathrm{Co-Co}}$ and $J_{\mathrm{Gd-Gd}}$, respectively). The resulting antiparallel configuration of $M_{\mathrm{\mathrm{net}}}$ corresponds to parallel alignment of the magnetization in the TM- and RE-dominant regions, and a retreat of the light blue DW in (c). The dark blue arrows in (a) and (c) represent the Co magnetization, which is imaged with MOKE microscopy.}\label{fig:Figschematic} 
\end{figure*}

To engineer spontaneous DW motion in our TM-RE ferrimagnet devices, we used direct-write laser annealing (DWLA) \cite{Riddiford2024} to create TM-dominant tracks within a RE-dominant Co$_{70}$Gd$_{30}$ film shown schematically in Fig.~\ref{fig:Figschematic}a (see Supporting Information Section S1 for details). The exposed tracks have different widths $w_{\mathrm{track}}$, defined in Fig.~\ref{fig:Figschematic}c, ranging from 500~nm to~7~\textmu m. To illustrate how this local patterning of ferrimagnetic properties yields self-driven DW motion in the track, we first apply an out-of-plane magnetic field \textit{H} as shown in Fig.~\ref{fig:Figschematic}a. If \textit{H} exceeds the coercive field but remains weaker than the antiferromagnetic Co-Gd coupling ($J_{\mathrm{Co-Gd}}$), the net magnetizations ($M_{\mathrm{net}}$) of all regions will align with \textit{H}. To highlight this, the detailed magnetic configuration across the green plane in Fig.~\ref{fig:Figschematic}a, consisting of two RE-dominant regions separated by a TM-dominant region, is shown in Fig.~\ref{fig:Figschematic}b. This parallel configuration of $M_{\mathrm{net}}$ means that the magnetization of the Co (Gd) sublattice alternates when going from left to right, pointing down-up-down (up-down-up). As a result of this alternating configuration, DWs are stabilized at the interface between regions with different dominant sublattices (light blue boundary in Fig.~\ref{fig:Figschematic}a). Unlike in conventional magnetic materials, these domain states are \textit{stabilized} by the applied field and are defined by a spatial variation in the sublattice magnetization orientation (as opposed to the local $M_{\mathrm{net}}$). 

When \textit{H} is removed (Fig.~\ref{fig:Figschematic}c), the impact of the ferromagnetic Co-Co and Gd-Gd exchange interactions ($J_{\mathrm{Co-Co}}$ and $J_{\mathrm{Gd-Gd}}$, respectively) emerges. Specifically, $J_{\mathrm{Co-Co}}$ and $J_{\mathrm{Gd-Gd}}$ favor parallel alignment of the Co and Gd sublattice magnetizations so that, when \textit{H} is removed, the lateral exchange coupling (LEC) across the boundaries between the TM- and RE-dominant regions promotes a spontaneous reversal of $M_{\mathrm{net}}$ in the TM-dominant region to minimize the ferromagnetic exchange energy as shown in Fig.~\ref{fig:Figschematic}d, corresponding to a cross-sectional view across the yellow plane in Fig.~\ref{fig:Figschematic}c. As we have shown previously, the LEC-induced magnetization reversal of a TM-dominant region with an apex-shaped end proceeds in a specific sequence \cite{Liu2023}: First, $M_{\mathrm{net}}$ is reversed within the apex of the TM-dominant region (Fig.~\ref{fig:Figschematic}c), where, as an interfacial interaction, the LEC strength is maximized. This reversal proceeds through motion of the DW that previously existed at the edges of the apex region (light blue boundary in Fig.~\ref{fig:Figschematic}a) into the TM-dominant region (Fig.~\ref{fig:Figschematic}c), observed as DW motion along the track. The curvature of the DW within the TM-dominant region (Fig.~\ref{fig:Figschematic}c) is a direct result of the fact that the LEC originates at the interface between the TM- and RE-dominant regions \cite{Liu2023}. This DW then propagates down the TM-dominant track, yielding parallel alignment of $M_{\mathrm{net}}$ across like sublattices (see Fig.~\ref{fig:Figschematic}d). 

While we previously cataloged the impacts of LEC through magnetometry measurements \cite{Liu2023}, we now determine the material factors that govern LEC-driven DW motion and how this DW motion can be coupled with spin-orbit torque to yield behaviors relevant for neuromorphic computing. We will show that the speed of LEC-driven DW motion directly determines the leak and reset behaviors essential to neuromorphic computing with magnetic DWs. Therefore, a precise understanding of how the speed of spontaneous DW motion can be tuned is critical for developing DW-based artificial neurons. Accordingly, we denote the velocity of self-driven DWs as $v_{\mathrm{leak/reset}}$ throughout this work. 

\begin{figure*}[htb]
    \centering
    \includegraphics[width=0.75\textwidth]{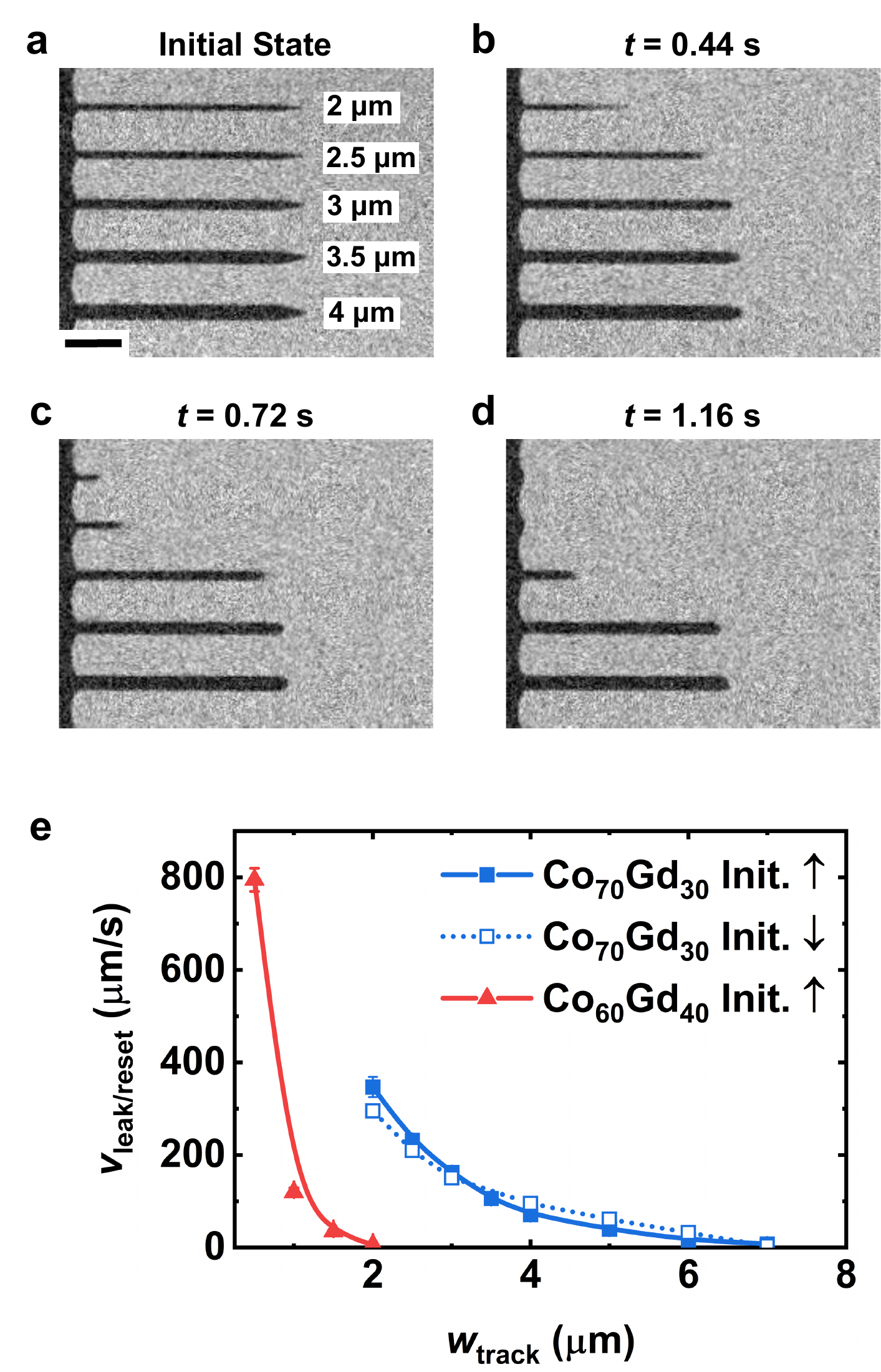}
    \caption{\textbf{Controlling self-driven DW leak and reset velocities by modifying the track width design and ferrimagnet composition.} \textbf{(a-d)} Polar MOKE images of the domain state in the Co$_{70}$Gd$_{30}$ sample while a +200~mT out-of-plane magnetic field was applied to initialize the sample (a), then at 0.44 s (b), 0.72 s (c), and 1.16 s (d) after the field was removed (scale bar = 15 \textmu m). $w_{\mathrm{track}}$ for each track is indicated in (a). In (a-d), black (grey) contrast corresponds to regions where the Co sublattice is magnetized up (down), corresponding to the TM-dominant (RE-dominant) regions. \textbf{(e)} The average DW leak and reset velocity $v_{\mathrm{leak/reset}}$ as a function of $w_{\mathrm{track}}$ is shown for the Co$_{70}$Gd$_{30}$ and Co$_{60}$Gd$_{40}$ samples (blue squares and red triangles, respectively). The lines are guides to the eye. For the Co$_{70}$Gd$_{30}$ sample, data for initialization in a -200~mT field is also provided (open blue squares). See Supporting Information Section~S2 for details on the determination of $v_{\mathrm{leak/reset}}$ and the error bars.}\label{fig:Fig1} 
\end{figure*}

To observe how $w_{\mathrm{track}}$ impacts $v_{\mathrm{leak/reset}}$, we began by initializing a parallel alignment of $M_{\mathrm{net}}$ across the TM- and RE-dominant regions using a +200~mT out-of-plane magnetic field as shown schematically in Fig.~\ref{fig:Figschematic}a. A magneto optic Kerr effect (MOKE) image of the magnetic configuration stabilized in this field for five tracks with different $w_{\mathrm{track}}$, ranging from 2 to 4 \textmu m is shown in Fig.~\ref{fig:Fig1}a. Because the white-light illumination we use for MOKE imaging is most sensitive to the Co sublattice \cite{Caretta2018,honda1988magneto,mangin2014engineered}, the contrast of the track (surrounding region) with the Co magnetization pointing out of (into) the plane appears dark (bright). After removing the magnetic field, we recorded MOKE images at 25~Hz (Supporting~Video~1). Within 0.44~s of removing the magnetic field, almost half of the track with $w_{\mathrm{track}}$~=~2~\textmu m has reversed, whereas smaller portions of the tracks with larger $w_{\mathrm{track}}$ have reversed (Fig.~\ref{fig:Fig1}b). 0.72~s and 1.16~s after the field was removed (Figs.~\ref{fig:Fig1}c and \ref{fig:Fig1}d, respectively), the DW displacements remain inversely related to $w_{\mathrm{track}}$. To quantify how $w_{\mathrm{track}}$ impacts the speed of LEC-propelled DWs, we determined $v_{\mathrm{leak/reset}}$ for the different $w_{\mathrm{track}}$ given by the solid blue squares in Fig.~\ref{fig:Fig1}e (see Supporting Information Section~S2 for details on the velocity calculations). We find that $v_{\mathrm{leak/reset}}$ decays with increasing $w_{\mathrm{track}}$, such that the LEC becomes too weak to promote spontaneous DW motion when $w_{\mathrm{track}}$~$\geq~$7~\textmu m. Furthermore, we find that this inverse relationship between $w_{\mathrm{track}}$ and $v_{\mathrm{leak/reset}}$ is the same for a reversed polarity of the initialization field (open blue squares in Fig.~\ref{fig:Fig1}e). The observed dependence of $v_{\mathrm{leak/reset}}$ on $w_{\mathrm{track}}$ aligns with prior observations that narrower tracks, being more interfacial in character, exhibit stronger LEC \cite{Liu2023}. However, we show for the first time that stronger LEC leads to faster self-driven DW motion.

While we see that $w_{\mathrm{track}}$ strongly influences the speed of self-driven DWs, it is also essential to determine how intrinsic material properties, such as the TM–RE ferrimagnet composition, affect LEC-driven DW motion. To accomplish this, we fabricated a Co$_{60}$Gd$_{40}$ sample with a 10~\% increase in Gd concentration (see Supporting Information Section~S1 for details). The Co$_{60}$Gd$_{40}$ sample was RE-dominant with perpendicular magnetic anisotropy in the as-grown state and received the same DWLA treatment as the Co$_{70}$Gd$_{30}$ sample. Using the same experimental procedure as above, we find that, while the value of $w_{\mathrm{track}}$ above which LEC cannot induce DW motion in the Co$_{70}$Gd$_{30}$ sample is 7~\textmu m, this value drops to 2~\textmu m in the Co$_{60}$Gd$_{40}$ sample (compare solid blue squares and red triangles in Fig.~\ref{fig:Fig1}e). In previous work, it was shown that a Co$_{60}$Gd$_{40}$ sample can have a net exchange stiffness $\sim$~60~\% lower than a Co$_{70}$Gd$_{30}$ sample \cite{suzuki2023compositional,gangulee,gangulee2}, which can result in weaker LEC, and thus, lower DW velocities (see Supporting Information Fig.~S1). In contrast, increasing the Gd content reduces the saturation magnetization of the as-grown film at room temperature by $\sim$~20~\% (see Figure~S1 in the Supporting Information), reducing dipolar interactions between the TM- and RE-dominant regions, potentially slowing LEC-induced DW motion -- a possibility we later demonstrate. While precisely distinguishing the relative impact of these two effects on the DW dynamics observed is nontrivial, Fig.~\ref{fig:Fig1}e nonetheless illustrates that changing the ferrimagnet composition provides another means of controlling LEC-driven DW motion. This understanding of how material properties affect LEC is crucial to the scalability of self-driven DW motion, as it allows the leak and reset dynamics of artificial neurons to be tuned for specific device sizes.

\begin{figure*}[htb]
    \centering
    \includegraphics[width=1\textwidth]{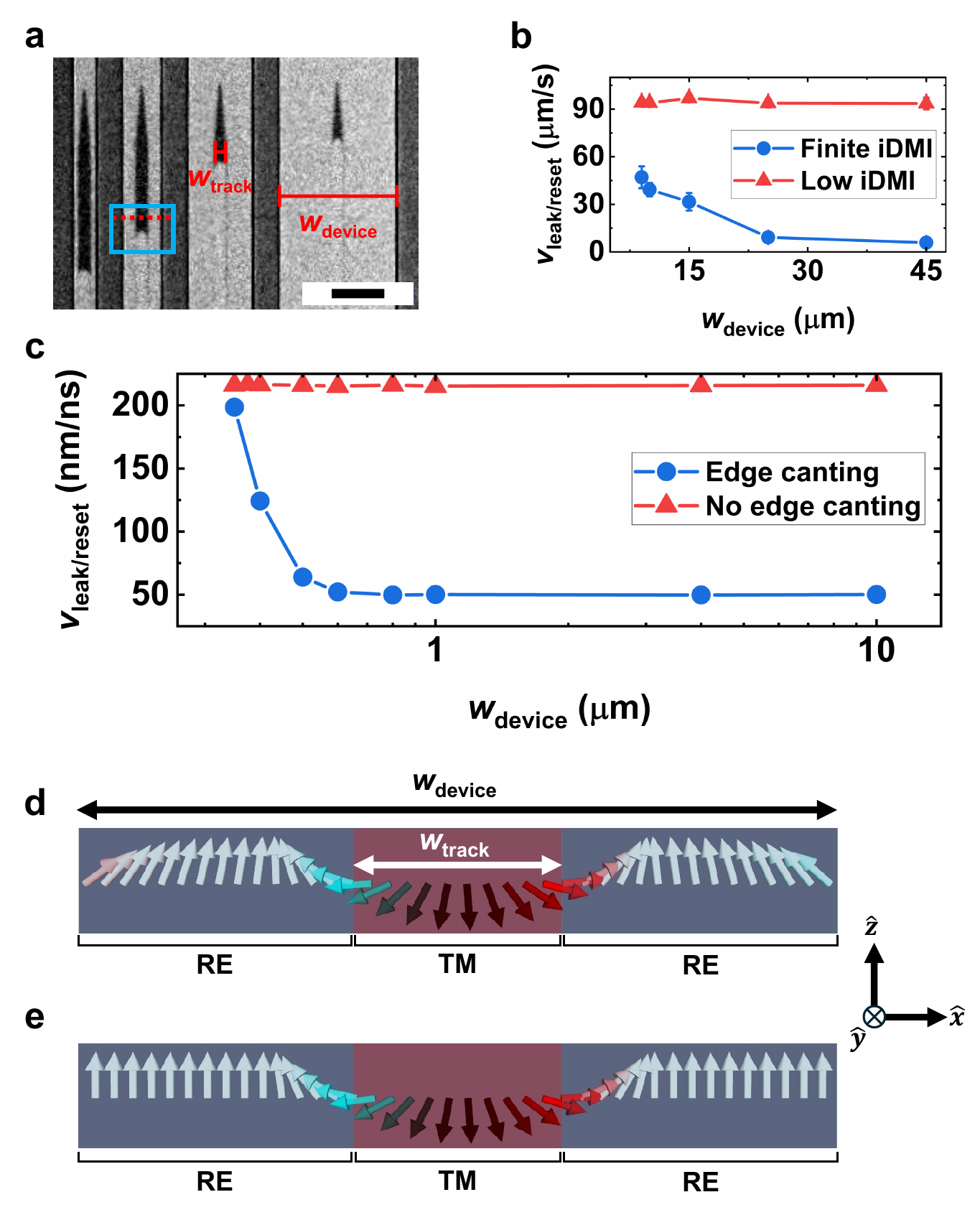}
    \caption{\textbf{Joint influence of device size and iDMI on DW leak and reset behaviors.} \textbf{(a)} Polar MOKE image of the domain state present in a Co$_{70}$Gd$_{30}$ sample with finite iDMI strength that has been lithographically patterned into devices of varying width $w_{\mathrm{device}}$, with $w_{\mathrm{device}}$~=~5~\textmu m,  2.52~s after the initializing magnetic field was removed. For this measurement, the background image subtracted from the displayed image was collected while applying a -200~mT magnetic field. (scale~bar~=~20~\textmu m). \textbf{(b)} The DW leak/reset velocity $v_{\mathrm{leak/reset}}$ as a function of $w_{\mathrm{track}}$ for the Co$_{70}$Gd$_{30}$ samples with finite and low iDMI. Details on how $v_{\mathrm{leak/reset}}$ was determined are provided in Section~S2 of the Supporting Information. \textbf{(c)} $v_{\mathrm{leak/reset}}$ as a function of $w_{\mathrm{track}}$ obtained from micromagnetic simulations for systems with and without chiral edge canting. \textbf{(d,e)} Cross-sectional views of the magnetic configuration along paths similar to the red dashed line in (a) for simulations with (d) and without (e) chiral edge canting. The simulations from which (d and e) were taken are provided in Figure~S4 of the Supporting Information.}
    \label{fig:Fig4} 
\end{figure*}

Since any potential technological application of magnetic DW motion inevitably requires maximizing areal density through device miniaturization, we now examine how LEC-driven DW motion is impacted by reducing the size of the RE-dominant region surrounding the TM-dominant tracks. To accomplish this, we prepared devices that feature TM-dominant tracks of fixed $w_{\mathrm{track}}$~=~5~\textmu m while varying the width of the surrounding device $w_{\mathrm{device}}$, defined in Fig.\ref{fig:Figschematic}c, from 9 to 45~\textmu m (Fig.~\ref{fig:Fig4}a and Supporting~Video~2). Repeating the same initialization procedure as before, we determined $v_{\mathrm{leak/reset}}$ as a function of $w_{\mathrm{device}}$ (blue circles in Fig.~\ref{fig:Fig4}b), finding that, for a fixed $w_{\mathrm{track}}$, there is a strong inverse relationship between $w_{\mathrm{device}}$ and $v_{\mathrm{leak/reset}}$. To understand the observed relationship between $w_{\mathrm{device}}$ and the measured $v_{\mathrm{leak/reset}}$, we micromagnetically simulated the LEC-induced DW dynamics. Full details on the simulation parameters are provided in the Supporting Information, including the interfacial Dzyaloshinskii-Moriya interaction (iDMI) present at the Pt/CoGd interface \cite{quessab2020tuning,blasing2018exchange,quessab2021interplay}. The simulated regions correspond to DW segments similar to the one enclosed by the blue box in Fig.~\ref{fig:Fig4}a. From these simulations, we extracted $v_{\mathrm{leak/reset}}$ as a function of $w_{\mathrm{device}}$ (blue circles in Fig.~\ref{fig:Fig4}c), finding a trend similar to the experimental results of Fig.~\ref{fig:Fig4}b. To determine the origins of this inverse relationship between $w_{\mathrm{device}}$ and $v_{\mathrm{leak/reset}}$, we consider the magnetic configuration of our devices 3 ns after the simulation began. Inspecting the cross-sectional view of the simulation with $w_{\mathrm{device}}$~=~400 nm shown in Fig.~\ref{fig:Fig4}d (corresponding to the dashed red line in Fig.~\ref{fig:Fig4}a), the magnetization rotates counterclockwise as one moves from the left to right across the DWs; this rotation is commensurate with the left-hand Néel-type chirality promoted by the iDMI at the Pt/CoGd interface \cite{quessab2020tuning,blasing2018exchange,quessab2021interplay}. Additionally, at the edges of the device, the magnetization cants away from the \textit{z}-axis, resulting in an additional magnetization component along the +\textit{x}-axis (-\textit{x}-axis) at the left (right) edge. The opposite canting at the left and right edges matches the DW chirality stabilized by the iDMI \cite{garcia2014nonreciprocal,yoo,rohart,pizzini}. 

Altogether, our simulations reveal that the chirality of both the DWs and the edge canting promoted by the iDMI dictate a slight yet discernible head-to-head arrangement of the magnetization between the edges of the device and the magnetization in the DWs. Such head-to-head configurations are known to incur a magnetostatic energy penalty~\cite{mcmichael1997head}, and one way to reduce the energy is for the DW to propagate through the structure, thereby eliminating the DW from the system. As $w_{\mathrm{device}}$ is reduced, and hence, the distance between the DWs and device edges is also lowered, the DW energy density increases, thus providing an additional impetus for DW motion beyond that provided by LEC alone (see Fig.~S4 in the Supporting Information). 

To confirm the role that DW chirality and edge canting play in modulating the dynamics of LEC-driven DWs in patterned devices, we repeated the simulations with the magnetization of the RE-dominant regions fixed out-of-plane, suppressing the chiral edge canting previously shown in Fig.~\ref{fig:Fig4}d. As a larger iDMI energy density is known to reduce the DW energy density \cite{je2013asymmetric}, this approach allows us to change the magnetic configuration stabilized while keeping the overall energy scale similar. Because freezing the magnetization of the RE-dominant region prevents the DW and edge magnetization profiles from assuming a head-to-head configuration (see the cross sectional view of a system with edge canting disabled in Fig.~\ref{fig:Fig4}e), the DW energy density becomes independent of $w_{\mathrm{device}}$ (see Fig.~S4 in the Supporting Information). As a result, there is no appreciable change in the simulated $v_{\mathrm{leak/reset}}$ with $w_{\mathrm{device}}$ when the RE-dominant region is frozen (red triangles in Fig.~\ref{fig:Fig4}c). To experimentally validate this link between the iDMI, $v_{\mathrm{leak/reset}}$, and $w_{\mathrm{device}}$, we fabricated a sample with weaker iDMI, and thus a weaker degree of edge canting, also finding no pronounced change in $v_{\mathrm{leak/reset}}$ with $w_{\mathrm{device}}$ as shown by the red triangles in Fig.~\ref{fig:Fig4}b (see Supporting Information Section~S1 for sample details). 

Having demonstrated how patterned track and device size, TM-RE composition, and iDMI affect the LEC-driven motion of DWs, we now apply this tunability to yield functionalities relevant to neuromorphic computing. Specifically, we now combine in a single device SOT-induced DW motion along the track with LEC-driven spontaneous DW motion in the opposite direction. To obtain SOT-induced DW motion, a charge current is passed through our devices. Here, the spin current generated via the spin-Hall effect in the Pt layer exerts SOT on the CoGd layer, which, coupled with the left-hand chiral Néel-type DWs stabilized by the iDMI originating from the Pt/CoGd interface, results in DW displacement parallel to the conventional current \cite{quessab2022zero,blasing2018exchange,quessab2021interplay}. 

\begin{figure*}[htb]
    \centering
    \includegraphics[width=1\textwidth]{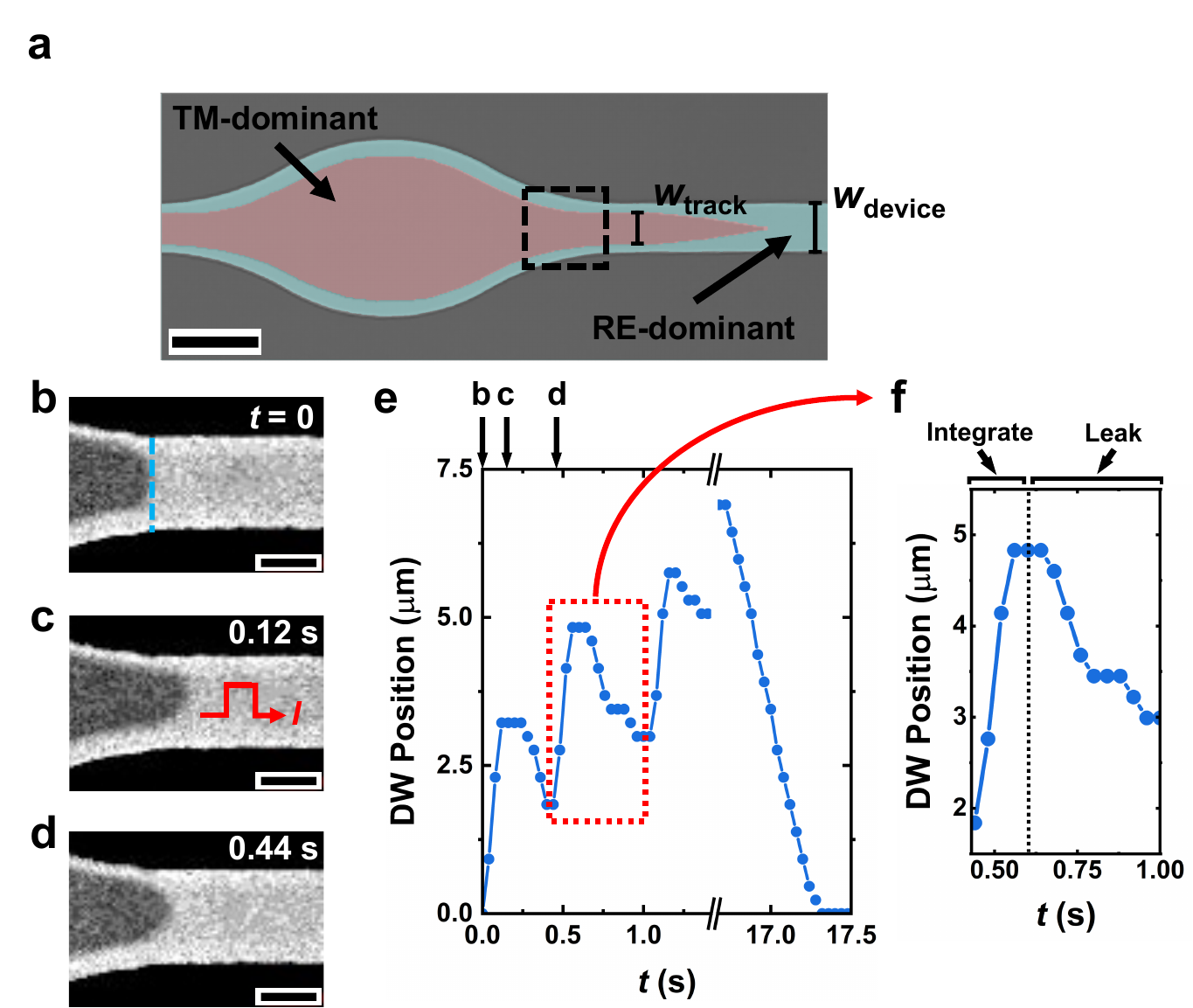}
    \caption{\textbf{Leaky integration and passive reset of a DW neuron.} \textbf{(a)} Optical microscopy image of the artificial neuron structure with a false overlay indicating the TM- and RE-dominant regions (scale bar = 10 \textmu m). \textbf{(b-d)} Polar MOKE images of the domain state present in a Co$_{70}$Gd$_{30}$ sample that has been magnetically and lithographically patterned to the specifications indicated in the main text (corresponding to the area in (a) enclosed by the black box) (b) after the initializing magnetic field was removed, (c) after the first current pulse was applied, and (d) immediately before the second current pulse was applied (scale bar = 5 \textmu m). The dashed box in (a) roughly indicates the area corresponding to the MOKE images in (b-d). The pulse sequence consisted of 120 \textmu s-long square pulses of current density 1 x 10$^{10}$ A/m$^{2}$ applied at a repetition rate of 2 Hz, with the direction of the conventional current flow \textit{I} indicated in (c). \textbf{(e)} The DW position as a function of time while current pulses were applied to the device, with the origin defined by the vertical dashed blue line in (b). The arrows show the times corresponding to the images in (b,c, and d). \textbf{(f)} A more detailed view of the leaky integration behavior of our DW neuron, corresponding to the data enclosed by the dashed red box in (e).}\label{fig:Fig2} 
\end{figure*}

To implement the leaky integration and reset functionalities in our materials, we used the device structure illustrated in Fig.~\ref{fig:Fig2}a. Leveraging our understanding of how $w_{\mathrm{track}}$ governs $v_{\mathrm{leak/reset}}$, we patterned a TM-dominant region with a spatially varying $w_{\mathrm{track}}$ within a lithographically-defined device designed to foster the current densities necessary for SOT-induced DW motion (Fig.~\ref{fig:Fig2}a). Specifically, the track contained an ellipse-shaped trap region where $w_{\mathrm{track}}$ > 7 \textmu m, thus \textit{inhibiting} spontaneous DW motion in response to LEC. In addition, within the trap region, $w_{\mathrm{device}}$ was expanded to minimize the current density within the trap, reducing the SOT below the threshold required to move the DW within the expanded portion of the device. Our device then operates as follows: After initializing the magnetic configuration in a +200~mT magnetic field, removing the magnetic field leads to LEC-induced magnetic reversal of the TM-dominant region up until $w_{\mathrm{track}}$ begins to expand, becoming too large to promote spontaneous DW motion, thus pinning the DW (Fig.~\ref{fig:Fig2}b). On applying an electrical current pulse from left to right, SOT displaces the DW to the right (Fig.~\ref{fig:Fig2}c). The electrical current pulse parameters are provided in the caption of Fig.~\ref{fig:Fig2}c. This SOT-induced DW motion emulates neuronal signal integration \cite{hassan2018magnetic,Sengupta_2018, can,siddiqui2019magnetic}. However, on moving rightward from the trap area, the DW moves to a region where $w_{\mathrm{track}}$~=~3.5~\textmu m, a value small enough to promote leftward LEC-induced motion. As such, during the interval between two SOT current pulses, the spontaneous DW motion back towards the trap area mimics a neuron leaking potential (Fig.~\ref{fig:Fig2}d) \cite{jaiswal2020neural,brigner2019graded,mah2021domain,mah2023leakage}. In the context of neuromorphic computing using magnetic DWs, neuron firing occurs when a DW reaches a defined position in the device \cite{hassan2018magnetic,Sengupta_2018, can,siddiqui2019magnetic}. In the present device, where LEC- and SOT-induced DW motion oppose each other, the firing behavior is determined by the balance between integration and leak dynamics. Notably, however, our material platform also supports integrate-and-fire behavior when LEC- and SOT-induced DW motion act cooperatively (see Section~S4 in the Supporting Information).

To better illustrate the leaky integration and passive reset behavior of our system under the joint influence of SOT and LEC, we plot the DW position as a function of time over many current pulses in Fig.~\ref{fig:Fig2}e (determined from the data shown in Supporting~Video~3). Defining the initial pinned position as the origin (blue dashed line in Fig.~\ref{fig:Fig2}b), Fig.~\ref{fig:Fig2}e provides a more detailed representation of the DW position than that shown in Figs.~\ref{fig:Fig2}b-d. Furthermore, once the current pulses are stopped (at \textit{t}~=~16.8~s), LEC returns the DW towards its original position, passively resetting the neuron potential. 

Having shown how the engineered presence of LEC enables the leaky integration and self-resetting functionalities essential for neuromorphic computing using magnetic DWs, we acknowledge that several desirable behaviors have yet to be realized in our systems. For example, the efficiency and accuracy of neural networks are greatly improved in systems exhibiting lateral inhibition, meaning that only the signal from the most responsive neuron is propagated forward through the network, while the output of competing neurons is suppressed (the winner-takes-all principle) \cite{hassan2018magnetic,coultrip1992cortical,cui2022intrinsic}. Furthermore, an ability to modify the interaction strength between neurons (the so-called synaptic weight) during operation is essential \cite{jin2019synaptic,siddiqui2019magnetic,liu2024domain}. These functionalities rely on the ability to locally and temporarily modulate the strength of LEC in our materials. For this, recent progress in the rapidly growing field of magnetoionics has shown that ionic gating can reversibly tune the magnetic properties of TM-RE ferrimagnets, including the exchange coupling strength and magnetically dominant sublattice \cite{huang2021voltage}, thus enabling on-demand, reversible control of LEC. These advances in materials control make lateral inhibition and synaptic weighting in ferrimagnet-based neuromorphic architectures increasingly within reach.

In conclusion, we have demonstrated how, for a neuron device consisting of a transition-metal dominant ferrimagnetic track coupled to a rare earth-dominant device region, the size and shape of magnetically patterned regions, composition of ferrimagnetic materials, and presence of chiral interactions enables regulation of the leak and reset rates of artificial neuron devices. By integrating control of spontaneous domain wall motion with spin-orbit torque, we have realized critical behaviors relevant to neuromorphic computing applications, including leaky integration and a passive reset of the domain wall position back to the initial state. Our device structure is easily implemented and scalable, leveraging materials grown via high-throughput, room-temperature sputter deposition, patterned through a single-step laser-induced process, and compatible with next-generation stimuli such as spin–orbit torque. Our findings therefore establish an important foundation for exploiting lateral exchange coupling effects for next-generation magnetic domain wall-based bioinspired computing architectures.

\section{Acknowledgements}
JAB acknowledges funding from the European Union’s Horizon 2020 research and innovation programme under the Marie Skłodowska-Curie grant agreement No 884104 (PSI-FELLOW-III-3i). AK acknowledges funding from the EU FET-Open RIA project SpinENGINE (Grant No. 861618). We thank the staff of the Laboratory for Nano and Quantum Technologies (LNQ) at PSI for cleanroom support. 

\section{Author Contributions}
JAB synthesized samples, performed experiments, analyzed the data, and conducted simulations. AK assisted with the sample preparation. JAB, AH, and LJH prepared the manuscript. All authors discussed the results and commented on the manuscript.








\pagebreak
\section*{Supporting Information}
\renewcommand\thesubsection{S\arabic{subsection}}
\setcounter{figure}{0}
\renewcommand{\thefigure}{S\arabic{figure}}
\subsection{S1 Sample Preparation}
\begin{figure*}[htb!]
    \centering
    \includegraphics[width=0.5\textwidth]{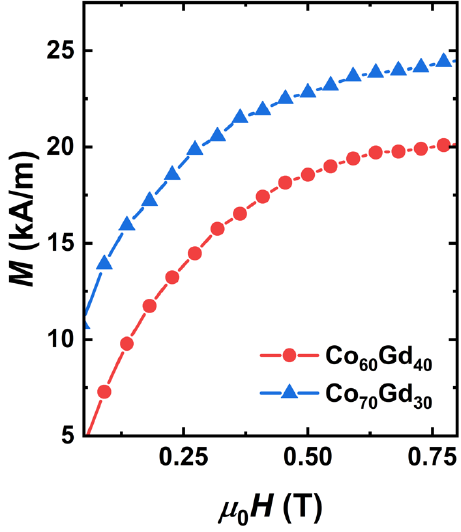}
    \caption{In-plane superconducting quantum interference device vibrating sample magnetometry (SQUID-VSM) measurements of the Co$_{70}$Gd$_{30}$ and Co$_{60}$Gd$_{40}$ samples (blue triangles and red squares, respectively). The error bars for the magnetization \textit{M} are smaller than the symbol size.} 
    \label{fig:FigS1}
\end{figure*}

Thin film heterostructures with the composition Ta (2 nm)/ Pt (5~nm)/ Co$_{x}$Gd$_{1-x}$~(6 nm)/ Ta (4 nm) were deposited onto thermally oxidized Si substrates (oxide~thickness~=~300~nm) by magnetron sputtering. All layers were sputtered in an Ar pressure of 0.4~Pa. The Ta layers were deposited using an RF power of 100 W, while Pt, Co, and Gd were deposited using DC powers of 100 W, 50 W, and 24 W, respectively. Deposition rates for each material were calibrated by growing reference samples for 10 minutes and measuring their thickness using x-ray reflectivity. The Co$_{x}$Gd$_{1-x}$ alloy layers were deposited by co-sputtering Co and Gd, with the stated atomic fractions determined by the sputtering powers used. In-plane superconducting quantum interference device vibrating sample magnetometry (SQUID-VSM) measurements indicate that the Co$_{70}$Gd$_{30}$ and Co$_{60}$Gd$_{40}$ films, each 6 nm thick, have saturation magnetizations of 25 kA/m and 20 kA/m, respectively (Fig.~\ref{fig:FigS1}). Both samples have an anisotropy field of approximately 0.75 T.

To determine the effect of lowering the iDMI energy density on LEC-induced DW motion, a Co$_{70}$Gd$_{30}$ sample of the composition Ti (4 nm)/ Co$_{70}$Gd$_{30}$ (8.5 nm)/ Ti~ (4 nm) was prepared, with the Ti deposited using a 100 W sputtering power. The thickness of the Co$_{70}$Gd$_{30}$ layer in this sample was made slightly thicker to achieve perpendicular magnetic anisotropy (PMA) through bulk anisotropy, as interfacial contributions to PMA are suppressed in this heavy-metal-free composition.

The dominant magnetic sublattice in our CoGd samples (which were RE-dominant in the as-grown state) was locally modified by direct-write laser annealing using a laser fluence of 1.4~J/cm$^{2}$. Full details on the direct-write laser annealing technique and its impact on the properties of TM-RE ferrimagnets can be found in Ref.~\cite{Riddiford2024}. The size of the RE-dominant region surrounding the TM-dominant tracks was controlled using UV photolithography and ion beam etching. A Heidelberg Instruments DWL 66+ system was used to pattern a mask in a layer of S1813 photoresist that was spin-coated onto the sample. After removing the exposed resist regions in a TMAH-based developer, an Ar$^{+}$ ion beam was used to etch the regions not covered by photoresist down to the substrate. 

\subsection{S2 MOKE Microscopy}
A magneto optic Kerr effect (MOKE) microscopy system manufactured by Evico Magnetics GmbH was used to collect images of DW motion in our devices. The microscope employs white LED illumination, which provides sensitivity to the orientation of the Co sublattice magnetization in our CoGd films \cite{Caretta2018,honda1988magneto,mangin2014engineered}. To improve image contrast, a reference image taken in a -200~mT out-of-plane magnetic field was subtracted from the recorded image before applying a +200 mT field to initialize the magnetic configuration of our devices. Any deviations from this protocol are noted in the main text. For each $w_{\mathrm{track}}$ or $w_{\mathrm{device}}$ considered, the initialization and measurement process was repeated five times, with the velocities averaged over the five trials. Error bars correspond to the standard error in average velocity over the five trials. Electrical current pulses were applied to the samples using an AVTECH AVR-E3-B-W1 pulse generator. The magnitude of the electrical current pulses supplied to the devices was monitored using an oscilloscope connected along the ground path.

\subsection{S3 Micromagnetic Simulations}
\begin{figure*}[htb!]
    \centering
    \includegraphics[width=1\textwidth]{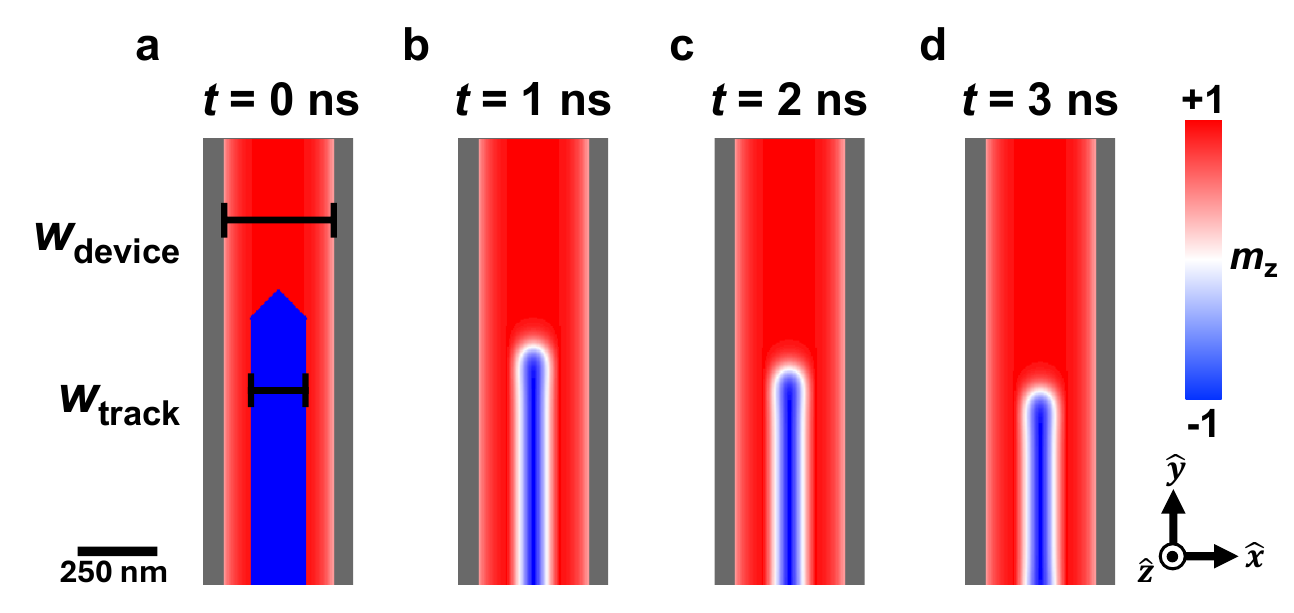}
    \caption{\textbf{(a)} Schematic of the normalized \textit{z}-component of the magnetization in the initialized state and \textbf{(b)} 1~ns, (\textbf{c}) 2~ns, and (\textbf{d}) 3~ns after the micromagnetic simulation began for a system with $w_{\mathrm{device}}$~=~400 nm, illustrating the simulated DW motion in response to LEC.} 
    \label{fig:FigS2}
\end{figure*}

Micromagnetic simulations were performed using the mumax$^{3}$ software package \cite{vansteenkiste2014design}. To model our materials, we have employed the following simulation parameters unless stated otherwise: Uniaxial perpendicular anisotropy constant $K$$_{u}$ = 1 x 10$^{4}$~J/m$^{3}$, exchange stiffness \textit{A}~=~15 pJ/m, saturation magnetization $M$$_{s}$~=~50~kA/m, interfacial Dzyaloshinskii-Moriya interaction (iDMI) energy density \textit{D}~=~+0.55 mJ/m$^{2}$ and ferrimagnet thickness \textit{t}~=~1~nm. The simulated devices had a length of 2 \textmu m and device widths $w_{\mathrm{device}}$ varying between 350 nm and 10 \textmu m. The simulation geometry was discretized into 5 nm x 5 nm x 1 nm cells (\textit{x}, \textit{y}, and \textit{z} directions, respectively). These coordinate axes are defined in Fig.~\ref{fig:FigS2}. The simulations were initialized by setting the magnetization of a 200 nm-wide region at the center of the simulation, representing the TM-dominant track, into the sample plane, while the magnetization in the surrounding areas was oriented out-of-plane (Fig.~\ref{fig:FigS2}a). Once the simulation began, LEC was sufficient to initiate domain wall motion starting from the apex of the simulated TM-dominant track (Fig.~\ref{fig:FigS2}b-d). Unlike the experimental results shown in the main text, our simulations demonstrate an initial decrease in $w_{\mathrm{track}}$ of the simulated TM-dominant track once the simulation begins. This behavior stems from the fact that in our simulations, the TM-dominant track and RE-dominant device have identical magnetic properties but opposing initial magnetic orientations. As such, our simulations do not account for DW pinning induced by the abrupt change in magnetic properties at the TM-RE boundary in our experimental devices \cite{van2016tunable,balavz2018static,franken2011domain}. However, by \textit{t}~=~1~ns, the width of the TM-dominant track stabilizes. Accordingly, in our analysis of the simulation results, we only consider the system when \textit{t} > 1~ns, where the TM-dominant region has reached a steady-state width.

\begin{figure*}[htb!]
    \centering
    \includegraphics[width=0.5\textwidth]{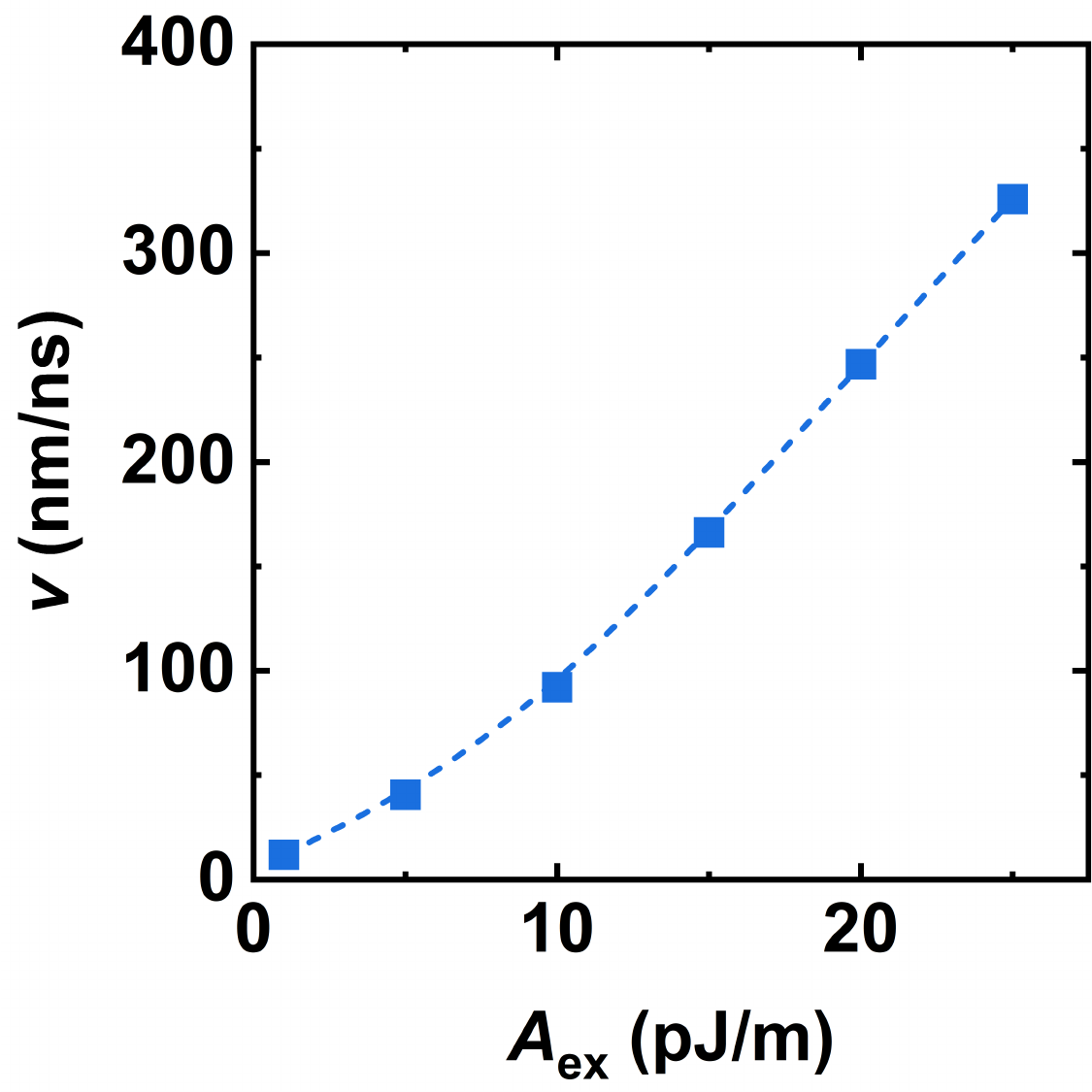}
    \caption{Simulated $v_{\mathrm{leak/reset}}$ as a function of exchange stiffness \textit{A}. The width of the simulated TM-dominant region was held constant at 200 nm. To simulate a semi-infinite film, periodic boundary conditions were applied along the \textit{x} and \textit{y} directions} 
    \label{fig:FigS3}
\end{figure*}

Micromagnetic simulations were initially used to determine the relationship between \textit{A} and the speed of DWs driven by LEC. Keeping all other magnetic parameters constant to the values stated above, \textit{A} was varied between 1 pJ/m and 25 pJ/m. Periodic boundary conditions were applied along the $\pm$\textit{x}-directions in the simulation to mimic a semi-infinite RE-dominant region surrounding the TM-dominant track.  We find that the DW leak and reset velocity ($v_{\mathrm{leak/reset}}$) exhibits a proportionality to \textit{A} (Fig.~\ref{fig:FigS3}), confirming that changing \textit{A} provides a material-intrinsic means of tuning the speed of DW motion in response to LEC, as shown in Fig.~2e of the main text. 

\begin{figure*}[htb!]
    \centering
    \includegraphics[width=0.75\textwidth]{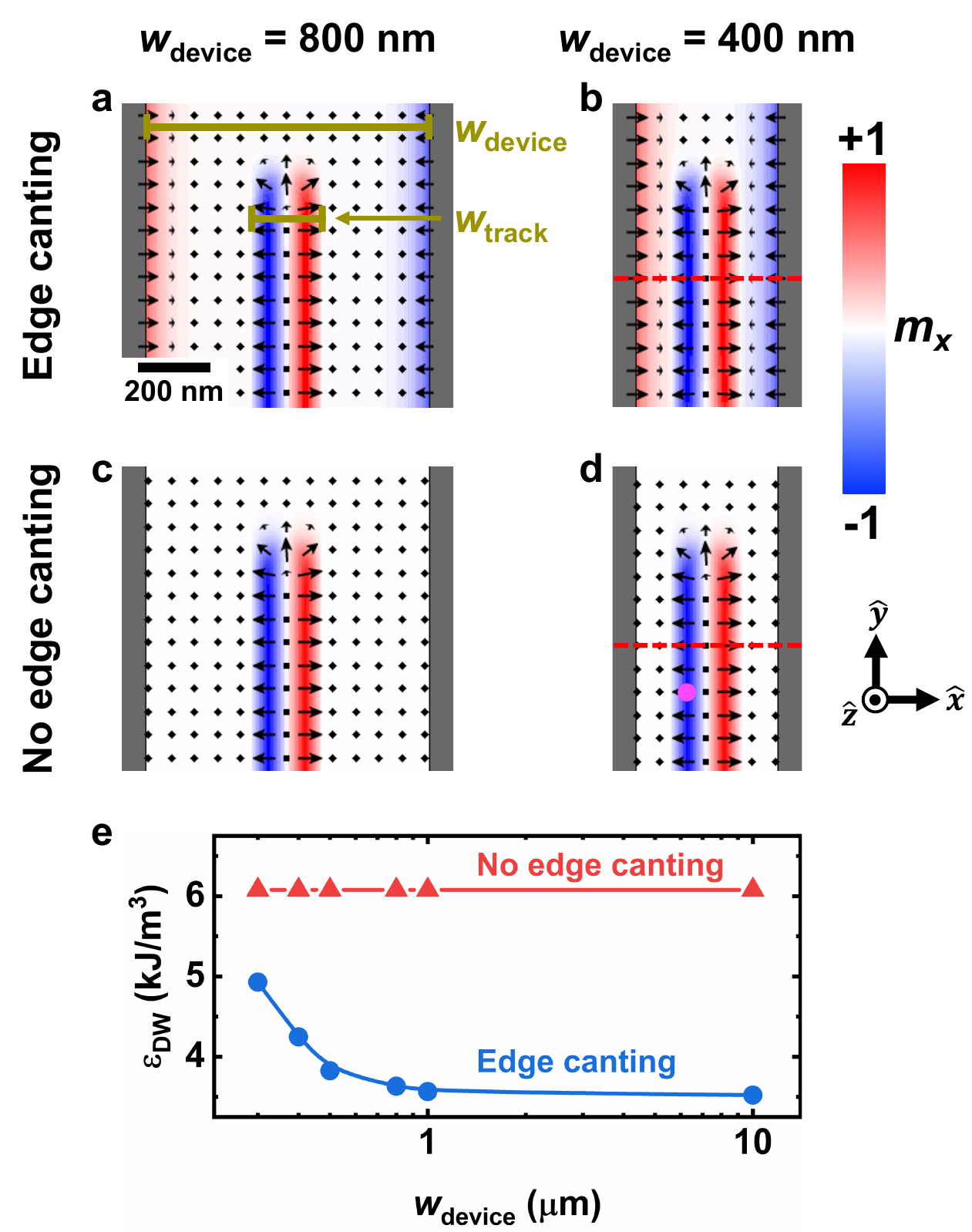}
    \caption{\textbf{(a-d)} Top-down depictions of the magnetic configuration of systems with (a,b) and without (c,d) chiral edge canting and $w_{\mathrm{track}}$ values of 800~nm (a,c) and 400~nm (b,d), 3~ns after the simulation began. The simulated regions shown correspond to regions similar to the one enclosed by the blue box in Fig.~3a of the main text. The cross sectional view shown in Fig.~3d (Fig.~3e) of the main text corresponds to the magnetization along the dashed red line in Fig.~S4b (Fig.~S4d)  \textbf{(e)} The total micromagnetic energy density $\varepsilon$$_{\mathrm{DW}}$ in the center of the DW region 3~ns after the simulation began for systems with (blue dots) and without (red triangles) edge canting for a variety of $w_{\mathrm{track}}$ values. The magenta dot in (d) indicates the position at which $\varepsilon$$_{\mathrm{DW}}$ was determined for each simulation. The lines connecting the symbols in (e) are guides to the eye.} 
    \label{fig:FigS4}
\end{figure*}

To micromagnetically determine how LEC-induced DW motion is impacted by the presence of the chiral edge canting promoted by the iDMI, we constrained the magnetization of the RE-dominant region surrounding the TM-dominant track to point along the +\textit{z}-axis. In Figs.~\ref{fig:FigS4}a,b (Figs.~\ref{fig:FigS4}c,d), we show the simulated magnetization profile of systems with chiral edge canting enabled (disabled) and $w_{\mathrm{device}}$ values of 800 nm and 400 nm, respectively, 3~ns after the simulation began. In contrast to when chiral edge canting is enabled (Figs.~\ref{fig:FigS4}a,b), a head-to-head arrangement between the magnetization of the DW and the edges of the device does not develop when chiral edge canting is disabled (Figs.~\ref{fig:FigS4}c,d), as further demonstrated by the cross-sectional view shown in Fig.~3e of the main text (corresponding to the dotted red line in Fig.~\ref{fig:FigS4}d). 

To understand how chiral edge canting of the magnetization affects the DW energy, and consequently the velocity of LEC-driven DWs, we examine the DW energy density $\varepsilon$$_{\mathrm{DW}}$ at the DW center in our simulations (location illustrated by the magenta dot in Fig.~\ref{fig:FigS4}d) 3~ns after the simulations began. When chiral edge canting is suppressed, $\varepsilon$$_{\mathrm{DW}}$ remains constant with respect to $w_{\mathrm{device}}$ (red triangles in Fig.~\ref{fig:FigS4}e). In contrast, when chiral edge canting and the associated partial head-to-head magnetic configuration are present (as in Fig.~3d of the main text), $\varepsilon$$_{\mathrm{DW}}$ decreases with increasing $w_{\mathrm{device}}$ (blue circles in Fig.~\ref{fig:FigS4}e). This trend reflects the fact that increasing $w_{\mathrm{device}}$ also enlarges the separation between the DW and the canted edge states, thereby weakening their dipolar interaction. Comparing Fig.~\ref{fig:FigS4}e with the DW velocity versus $w_{\mathrm{device}}$ data shown in Fig.~3c of the main text, the higher DW velocity at smaller $w_{\mathrm{device}}$ can be understood as follows: Because DWs with higher energy densities are less energetically favorable and LEC-induced reversal of the TM-dominant region removes the DW and its associated energy from the system, it is more energetically favorable for the TM-dominant region to switch in the case of narrower $w_{\mathrm{device}}$, leading to faster LEC-driven DW motion.

\subsection{S4 Alternate pathway to integrate-and-fire behaviors}
\begin{figure*}[htb!]
    \centering
    \includegraphics[width=0.85\textwidth]{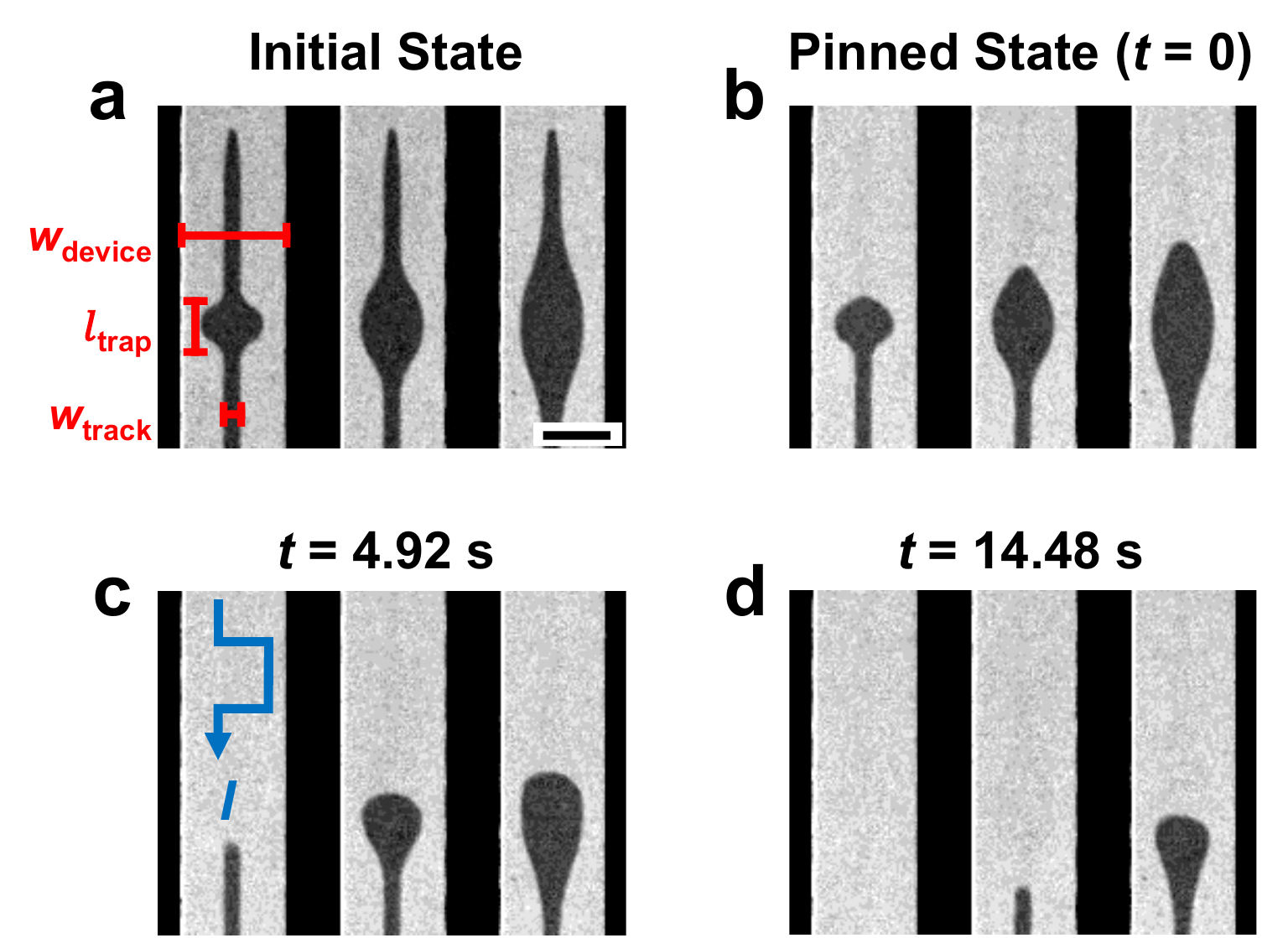}
    \caption{\textbf{(a)} Polar MOKE images of the domain state present in a Co$_{70}$Gd$_{30}$ sample engineered to have TM-dominant strips with elliptical trap regions with trap lengths $l_{trap}$ of 20 \textmu m, 40 \textmu m, and 70 \textmu m (left to right, respectively) while a +200 mT perpendicular magnetic field was applied to initialize the sample (scale bar = 20 \textmu m). \textbf{(b)} The magnetic configuration some time after the initializing magnetic field was removed. The domain configuration (\textbf{c}) 4.92 s and (\textbf{d}) 14.48 s after electrical current pulses were applied to the sample. The pulse sequence consisted of 120 \textmu s-long square pulses of current density 1 x 10$^{10}$ A/m$^{2}$ applied at a repetition rate of 2 Hz, with the direction of the conventional current flow \textit{I} indicated in (c).}\label{fig:FigS5} 
\end{figure*}

In addition to the leaky integration and passive reset behaviors demonstrated in Fig.~4 of the main text, which arise when LEC- and SOT-driven DW motion oppose each other, our materials, devices, and stimuli can be tuned to produce integrate-and-fire functionality when LEC and SOT drive DWs in the same direction. Here, we demonstrate how the length of a trap region ($l_{trap}$), where $w_{\mathrm{track}}$ is wide enough to prevent spontaneous DW motion, can be used to adjust the integration behavior of DW-based neuronal structures, using the devices shown in Fig.~\ref{fig:FigS5}a. These devices feature ellipse-shaped trap regions where $w_{\mathrm{track}}$ is greater than 7 \textmu m, so as to prevent LEC-driven DW motion in response to LEC. The design principle of these devices is that, within the trap regions, displacement of a DW solely in response to the SOT generated by the current pulses emulates a neuron integrating signal. Varying $l_{trap}$ from 20~\textmu m to 40~\textmu m to 70~\textmu m (from left to right in Fig.~\ref{fig:FigS5}a) provides a means to adjust the DW displacement that must be integrated before the DW exits the trap regions. Outside the traps, $w_{\mathrm{track}}$ was fixed to 4.5~\textmu m, a value small enough to ensure spontaneous LEC-driven DW motion. This allows the domain wall, once displaced from the trap by SOT, to spontaneously propagate to the end of the track, where its arrival signifies neuronal firing.\cite{hassan2018magnetic,Sengupta_2018,cui2022intrinsic}.

We now demonstrate these integration and firing functions, first initializing the net magnetization of the TM- and RE-dominant regions in a parallel state by applying a +200 mT perpendicular magnetic field to the device (Fig.~\ref{fig:FigS5}a). Upon removing the magnetic field, magnetization reversal of the upper part of the TM-dominant region of the track proceeds via DW motion, with the domain wall nucleated in the apex region and subsequently moving through the device until becoming pinned at the top ends of the traps (Fig.~\ref{fig:FigS5}b). Next, current pulses were applied simultaneously to all three track structures, enabling concurrent imaging of how the DWs move through the traps in response to the same current pulse amplitude. Detailed information on the current pulse sequence used is provided in the caption of Fig.~\ref{fig:FigS5} and the direction of the conventional current flow \textit{I} is indicated in Fig.~\ref{fig:FigS5}c. 

After applying the current pulses for 4.92~s, the gradual accumulation of DW displacement over the course of several current pulses mimics the neuronal integration process (Fig.~\ref{fig:FigS5}c). When the DW is fully displaced from the trap and into a region where $w_{\mathrm{track}}$ is low enough to promote spontaneous DW motion, the DW rapidly propagates to the end of the track - a response that corresponds to firing of the artificial neuron upon reaching a threshold potential. In the devices with $l_{trap}$~~$>$~20~\textmu m, SOT alone is insufficient to displace the DWs within the same time interval, and the DWs remain confined to the portion of the track where $w_{\mathrm{track}}$ is larger. As a result, the devices with longer trap regions require more DW displacement before they fire (Fig.~\ref{fig:FigS5}d). As such, for a given stimulus, the integration required for a DW to traverse the trap is intimately linked to our ability to locally control the action of LEC through $w_{\mathrm{track}}$. Furthermore, as discussed in the main text, magnetoionic tuning of the exchange coupling strength could be used as reconfigurable way to adjust the integration threshold for DW neuron firing without changing the size of the TM-dominant track.

\section{Supporting Video Captions}
 
\textbf{Supporting Video 1:} Polar MOKE video of the domain evolution of the Co$_{70}$Gd$_{30}$ sample discussed in Fig.~2 of the main text. The video begins while a +200 mT out-of-plane magnetic field was applied to initialize the sample and tracks the evolution in the position of the DWs along the tracks after the field was set to zero. Details on the design characteristics are provided in the main text.
\newline
\newline
\textbf{Supporting Video 2:} Polar MOKE video of the domain state present in the device shown in Fig.~3a of the main text. The video starts while a +200 mT out-of-plane magnetic field was applied to the sample and shows the evolution in the position of the DWs along the tracks once this field was removed.  
\newline
\newline
\textbf{Supporting Video 3:} Polar MOKE video of the  leaky integration and passive reset behavior of the neuronal structure shown in Fig.~4 of the main text as SOT current pulses are applied. The video begins after the initializing magnetic field has been removed and LEC-driven domain wall motion has occurred, resulting in the DW becoming pinned on the left side of the image where $w_{\mathrm{track}}$ becomes large enough to suppress further spontaneous DW motion. As electrical current pulses are applied, SOT-induced DW motion to the right corresponds to the integration of the input signal, whereas the LEC-driven DW motion to the left between SOT pulses mimics neuronal leaking. When the SOT current pulses are stopped 16.8~s into the video, LEC causes the DW to retreat to the left, towards the original position of the DW. Details on the SOT pulse characteristics are provided in the caption of Fig.~4 of the main text.



\bibliography{refs}
\end{document}